\documentclass[]{aa}
\usepackage{psfig}
\usepackage{txfonts}
\usepackage{graphicx}
\usepackage{amssymb}
\usepackage[below]{placeins}
\usepackage{natbib}
\bibpunct{(}{)}{;}{a}{}{,}

\def \arcmin{^{\prime}}

\begin{document}

\title{XMM-Newton high-resolution spectroscopy reveals the chemical evolution of M~87}
\author{N. Werner\inst{1}
 \and H. B{\"o}hringer\inst{3}
 \and J.S. Kaastra\inst{1}
 \and J. de Plaa\inst{1,2}
 \and A. Simionescu\inst{3}
 \and Jacco Vink\inst{2}}

\offprints{N. Werner, email {\tt n.werner@sron.nl}}

\institute{     SRON Netherlands Institute for Space Research, Sorbonnelaan 2,
                NL - 3584 CA Utrecht, the Netherlands
         \and
                Astronomical Institute, Utrecht University, P.O. Box 80000,
                NL - 3508 TA Utrecht, the Netherlands 
	\and 
		Max-Planck-Institut f{\"u}r extraterrestrische Physik, 85748 Garching, Germany}

\date{Received, accepted }

\abstract{We present here a study of chemical abundances in the giant elliptical galaxy M~87 using high-resolution spectra obtained with the Reflection Grating Spectrometers during two deep XMM-Newton observations. 
While we confirm the two-temperature structure of the inter-stellar medium (ISM) in M~87, we also show that a continuous temperature distribution describes the data equally well. 
The high statistics allows us, for the first time, to determine relatively accurate abundance values also for carbon and nitrogen. The comparison of the abundance ratios of C, N, O and Fe in the ISM of M~87 with those in the stellar population of our Galaxy shows that the relative contribution of core-collapse supernovae to the enrichment of the ISM in M~87 is significantly less than in the Milky Way and indicates that the enrichment of the ISM by iron through Type~Ia supernovae and by carbon and nitrogen is occurring in parallel.
This suggests that the main source of carbon and nitrogen in M~87 are the low- and intermediate-mass asymptotic giant branch stars. From the oxygen to iron abundance ratio in the hot gas
we estimate that the relative number of core collapse and type Ia supernovae contributing to the enrichment of the ISM in the core of M~87 is $\sim60$\% and $\sim40$\% respectively. The spatial distributions of iron and oxygen are different. While the oxygen abundance distribution is flat the iron abundance peaks in the core and has a gradient throughout the $4\arcmin$ wide field of view of the instrument, suggesting an early enrichment by core-collapse supernovae and a continuous contribution of type Ia supernovae.

}  

\maketitle

\section{Introduction}
\label{intro}

The giant elliptical galaxy M~87 resides at the center of the nearby Virgo Cluster. Its relatively small distance of $\sim16$~Mpc makes it a perfect target to study 
processes that are important in all clusters of galaxies, but to a much greater detail than possible elsewhere. Among such studies, the investigation of the structure and physics of cooling cores (previously thought to harbour cooling flows) and of the heavy element enrichment of the intra-cluster medium (ICM)  gain a central importance. This is due to the fact that M~87 is on one hand a famous cooling core cluster and on the other hand provides with an ICM temperature range of about $1-3$~keV an X-ray spectrum very rich in metal emission lines that allows us to perform detailed diagnostics of the chemical ICM abundances.  Consequently, M~87 is among the best studied objects in the sky.  Studies of M~87 at all wavelengths offer a detailed insight into the balance of the ICM cooling and the heating by the central active galactic nucleus (AGN). 

The radio map of M~87 shows two lobes of synchrotron emission located about $20-30$~kpc East and South-West of the nucleus \citep[]{owen2000}. The AGN and its inner jet probably supply the lobes with
energy and relativistic electrons. 
Based on the previous XMM-Newton observation \citet{belsole2001} and \citet{matsushita2002} found that the ICM in M~87 has locally a single phase structure, except for the regions
associated with radio lobes, where the X-ray emission can be modelled with two thermal components with temperatures of about 2 and 1~keV. The cooler gas fills a small volume
compared to the hotter component and is confined to the radio arms, rather than being associated with the potential well of the galaxy \citep{molendi2002}. Based on radio and
X-ray observations \citet{churazov2001} developed a model, according to which the bubbles seen in radio buoyantly rise through the hot plasma in the cluster and during their
rise they uplift cooler X-ray emitting gas from the central region out to larger distances. 
This scenario was also studied in numerical hydrodynamical simulations by \citet{bruggen2002} who illustrated the turbulent mixing of hot, cooler and relativistic plasma initiated by the bubble-ICM interaction. This effect probably gives rise to the observed two-phase or multi-phase temperature structure in the interaction region. Plasma cooling below a temperature of 0.8~keV is not observed in the XMM Reflection Grating Spectrometers (RGS) X-ray spectra \citep{sakelliou2002} and neither in the XMM European Photon Imaging Camera (EPIC) spectra of M~87 \citep{bohringer2002}. 

\citet{bohringer2001} and \citet{finoguenov2002} showed that oxygen has a relatively flat radial
abundance distribution compared to the steep gradients of silicon, sulfur, argon, calcium and iron. The strong abundance increase of iron in the core of M~87 compared to that
of oxygen indicates an enhanced contribution of type Ia supernovae (SN~Ia) in the central regions of the galaxy. The flatter oxygen gradient was also confirmed by
\citet{matsushita2003}.  On the other hand, while \citet{gastaldello2002} confirm the strong heavy-element gradient they also find a significant increase of abundance for light
elements, in particular oxygen, toward the center of the galaxy which contradicts the results of \citet{bohringer2001}, \citet{finoguenov2002} and \citet{matsushita2003}. 

In this paper we report the results obtained by combining two deep observation with the Reflection Grating Spectrometer \citep[RGS,][]{herder2001} aboard XMM-Newton. The strongly peaked high X-ray luminosity makes M~87 an ideal target for the RGS.  The high statistics allows us for the first time to reliably determine the abundance of carbon and nitrogen. 

While it is well known that elements from oxygen up to the iron group are primarily produced in supernovae, the main sources of carbon and nitrogen are still being debated. Both elements are believed to originate from a wide range of sources including winds of short-lived massive metal rich stars, longer-lived low- and intermediate-mass stars and also an early generation of massive stars \citep[e.g.][]{gustafsson1999,chiappini2003,meynet2002}. 
\citet{shi2002} find that carbon is enriched by winds of metal-rich massive stars at the beginning of the Galactic disk evolution, while at the later stage it is produced mainly by low-mass stars. 
They also conclude that nitrogen is produced mainly by intermediate-mass stars. 
\citet{bensby2006} find that the carbon enrichment in the Galaxy is happening on a time scale very similar to that of iron. They conclude that while at low metallicities the main carbon contributors are massive stars, carbon is later produced mainly by asymptotic giant branch (AGB) stars.

The deep observation also allows us to extract spatially resolved high-resolution spectra and to determine radial temperature profiles and abundance profiles of several elements out
to $3.5\arcmin$ from the core of M~87. 

For the distance to M~87 we adopt a value of 16.1~Mpc \citep{tonry2001} which implies a linear scale of 4.7~kpc~arcmin$^{-1}$. 
Unless specified otherwise, all errors are at the 68\% confidence level for one interesting parameter ($\Delta \chi^{2}=1$).

\section{Observations and data reduction}

M~87 was observed with XMM-Newton on June 19 2000 for 60~ks and re-observed in January 2005 with an exposure time of 109 ks.
We process the RGS data from both observations with the 6.5.0 version of the XMM-Newton Science Analysis System (SAS) following the method described in \citet{tamura2001b}. 

In order to minimise the effect of soft protons in our spectral analysis, we extract a light-curve using the events on CCD~9 of the RGS, outside the central area, with a distance larger than 30$\arcsec$ from the dispersion axis and
cut out all time intervals where the count rate deviates from the mean by more than $3\sigma$. 
After removing the high background periods from the data, the net effective exposure time is 40~ks for the first observation and 84~ks for the second observation. 

Because the extended emission fills the entire field-of-view of the RGS we model the background using the standard RGS background model available in SAS. We correct the standard response for the RGS instruments by additional, time dependent large scale effective area corrections found by one of us (Jacco Vink;
more information can be found on the website of the XMM-Newton users group\footnote{http://xmm.vilspa.esa.es/external/xmm\_user\_support/usersgroup/\\20060518/rgs\_calib\_eff.pdf}), which are based on the standard XMM-Newton calibration sources MRK~421 and PKS~2155-304. These corrections will be implemented
in the forthcoming release of the XMM-Newton SAS software package and calibration data files.

In order to derive spatial temperature and abundance profiles from the RGS we extract spectra from four $1\arcmin$ wide regions along the cross-dispersion axis of the instrument. We select the events from rectangular areas on the CCD strip in the cross-dispersion direction (see Fig~\ref{strips}).

The observed line emission of the hot gas is broadened by the spatial extent of the source along the dispersion direction. In order to account for this effect in the spectral modelling we convolve the model with the surface brightness profile of the galaxy along the dispersion direction \citep{tamura2004}. We derive the surface brightness profile for each extraction region from the EPIC/MOS1 image in the $0.8-1.4$~keV band along the dispersion direction of the RGS and convolve it with the RGS response in order to produce a predicted line spread function
(lsf). Because the radial profile for an ion can be different from this intensity profile, this method is not ideal. Therefore, we introduce two additional parameters, the width and the centroid of the lsf which are left free during spectral fitting in order to match the observed profiles of the main emission lines. The scale parameter $s$ for the width is the ratio of the observed lsf width and the width of the intensity profile.

\section{Spectral models}
\label{models}

For the spectral analysis we use the SPEX package \citep{kaastra1996}. 
We model the Galactic absorption using the \emph{hot} model of that package, which calculates
the transmission of a plasma in collisional ionization equilibrium (CIE) with cosmic abundances. We mimic
the transmission of a neutral gas by putting its temperature to 0.5 eV. Compared to the older standard absorption models like \citet{morrison}, the \emph{hot} model is using continuum opacities from \citet{verner1995}. Line opacities for most ions are from \citet{verner1996}.
To find the best description of the emission of the hot plasma in M~87 we use a combination of two
CIE plasma models (MEKAL) with coupled abundances. We also try to model the emission of the hot plasma with a differential emission measure
(DEM) model with a cut-off power-law distribution of emission measures versus temperature ({\it{wdem}}). The {\it{wdem}} model appears to be a good empirical approximation for
the spectrum in cooling cores of clusters of galaxies \citep[e.g.][]{kaastra2004,werner2006,deplaa2006}. 
The emission measure $Y = \int n_{\mathrm{e}} n_{\mathrm{H}} dV$ (where $n_{\mathrm{e}}$ and $n_{\mathrm{H}}$ are the electron and proton densities, $V$ is the volume of the
source) in the {\it{wdem}} model is specified in Eq.~(\ref{eq:wdem}) adapted from \citet{kaastra2004}:
\begin{equation}
\frac{dY}{dT} = \left\{ \begin{array}{ll}
AT^{1/\alpha} & \hspace{1.0cm} T_{\mathrm{min}} < T < T_{\mathrm{max}}, \\
0 & \hspace{1.0cm} \mathrm{elsewhere}. \\
\end{array} \right.
\label{eq:wdem}
\end{equation}

The emission measure distribution has a cut-off at $T_{\mathrm{min}}=cT_{\mathrm{max}}$.
For $\alpha \to \infty$ we obtain a flat emission measure distribution. 
The emission measure weighted mean temperature $T_{\mathrm{mean}}$ is given by:
\begin{equation}
T_{\mathrm{mean}}=\frac{\int^{T_{\mathrm{max}}}_{T_{\mathrm{min}}} \frac{dY}{dT}\, T\, dT}{\int^{T_{\mathrm{max}}}_{T_{\mathrm{min}}} \frac{dY}{dT}dT}.
\end{equation}
By integrating this equation between $T_{\mathrm{min}}$ and $T_{\mathrm{max}}$ we obtain a direct relation between $T_{\mathrm{mean}}$ and $T_{\mathrm{max}}$ as a function of
$\alpha$ and $c$:
\begin{equation}
T_{\mathrm{mean}}=\frac{(1+1/\alpha)(1-c^{1/\alpha+2})}{(2+1/\alpha)(1-c^{1/\alpha+1})}T_\mathrm{max}.
\end{equation}
A comparison of the \emph{wdem} model with the classical cooling-flow model can be found in \citet{deplaa2005a}. We note that the \emph{wdem} model contains less cool gas than
the classical cooling-flow model, which is consistent with recent observations \citep[]{peterson2001,peterson2003}.

The spectral lines in the MEKAL model are fitted self consistently. From the fits we obtain the abundances of all elements with detected line emission.
Throughout the paper we give the measured abundances relative to the proto-solar values given by \citet{lodders2003}.
We note that the values of the fitted elemental abundances do not depend on the chosen values for the solar abundances. We give the values of the elemental abundances with
respect to solar for convenience. The recent solar abundance determinations by \citet{lodders2003}
give significantly lower abundances of oxygen, neon and iron than those measured by \citet{anders1989}.

\subsection{The AGN spectrum}

The center of M~87 harbours an active galactic nucleus (AGN). Its emission can be well fitted with a power-law. The good description of the AGN emission is essential for the determination of the absolute abundances in the core of M~87 with the RGS. In the first observation we determine the parameter values of the power-law  by fitting the EPIC spectrum of the core of the galaxy. We extract the spectrum from a small circular region with a radius of 20\arcsec\ and fit it with a power-law and two thermal models. The parameter values of the power-law strongly depend on the value for the absorption column density in the model. For the best measured value of the absorption column density $N_{\mathrm H}=1.8\times10^{20}$~cm$^{-2}$ \citep{lieu1996a} we obtain a power-law with a photon index of $\Gamma=2.18\pm0.03$ and a $2-10$ keV luminosity of ($8.6\pm0.5$)$\times10^{40}$  erg\, s$^{-1}$. 

The AGN was much brighter during the second observation than during the first one and also its photon index changed. 
However, the strong pile-up both in EPIC/MOS and EPIC/pn does not allow us to determine the properties of the AGN emission from the EPIC observations. We have to do it using the RGS which is much less susceptible to pile-up. 
Since the two observations were performed with similar position angles the thermal cluster emission in the two datasets is the same. We determine the parameter values of the cluster emission in the first dataset. 
While we fit the parameter values for the power-law emission in the second dataset we keep the parameter values of the cluster emission fixed. We find a power-law with a photon index of $\Gamma=1.95\pm0.03$ and a $2-10$ keV luminosity of $(1.15\pm0.11)\times10^{42}$  erg\, s$^{-1}$. 
To confirm this high flux, we scale the RGS spectrum obtained during the first observation to the exposure time of the second observation and we subtract it from the RGS spectrum from the second dataset. This way we get rid of most of the cluster emission, which allows us to analyze the excess power-law emission from the central AGN detected during the second observation. With this exercise we confirm our previous results about the high AGN luminosity in the second dataset. 

While fitting the thermal ICM/ISM emission we keep the power-law parameters fixed at the best determined values. 
The uncertainties in the AGN luminosity introduce systematic uncertainties in the emission measure of the thermal components and in the absolute abundance values. The uncertainties in the photon index of the power-law emission are much smaller than those in the luminosity and thus have a smaller influence on the best fit parameters of the thermal components. If the luminosity of the AGN in the second observation is 10\% lower than the adopted value, then the best fit emission measures of the thermal components will be $\sim15$\% higher and the absolute abundance values will be $\sim15$\% lower. If the luminosity of the AGN is 10\% higher than the adopted value, then the best fit emission measures will be $\sim15$\% lower and the absolute abundance values $\sim15$\% higher. The best fit temperatures and relative abundance values are, however, robust and the systematic uncertainties arising from the uncertainty in the AGN emission are smaller than their statistical errors.

\section{Results}

\subsection{The combined dataset}

\begin{figure}
\includegraphics[width=6.7cm,clip=t,angle=270.]{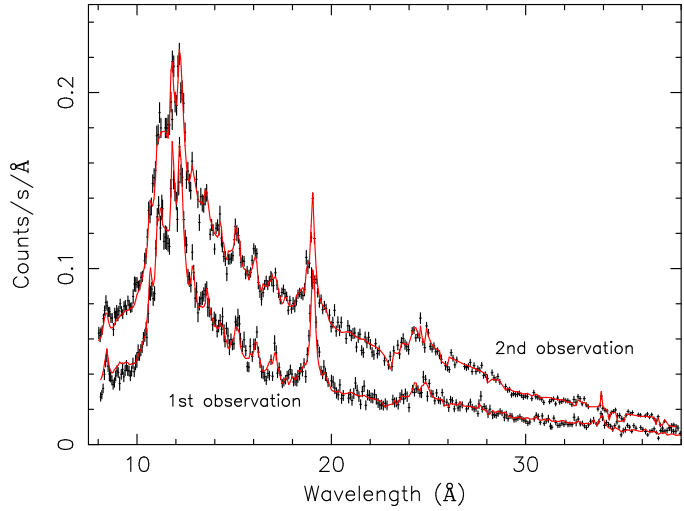}
\caption{1st order RGS spectra obtained during two observations with different luminosities of the central AGN. The two spectra are extracted from a $1.1\arcmin$ wide stripe centred on the core of M~87 and are fitted simultaneously. The total useful exposure time is 124~ks. The continuous line represents the best fit model of the spectrum. }
\label{total}
\end{figure}

\begin{figure}
\includegraphics[width=6.7cm,clip=t,angle=90.]{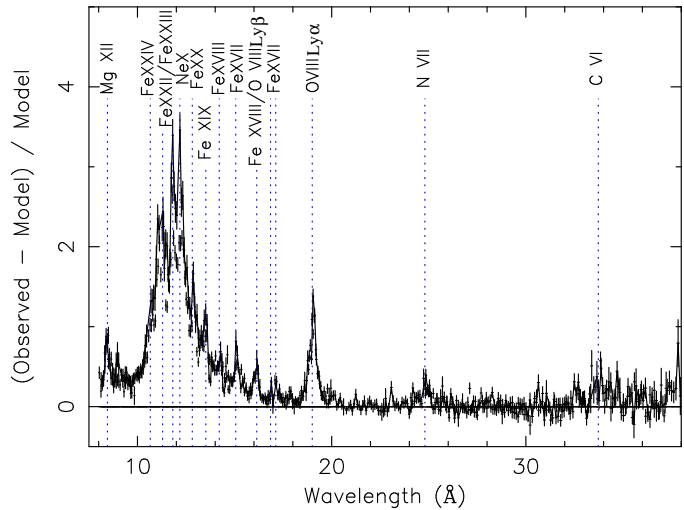}
\caption{Residuals of the fit to the combined RGS spectrum with a total exposure time of 124~ks with line emission set to zero in the
model. We indicate the positions of all the detected spectral lines.}
\label{lines}
\end{figure} 

\begin{table}
\caption{Fit results for the RGS spectra extracted from a 1.1$\arcmin$ wide strip centred on the core of M~87. 
The iron abundance is given with respect to the proto-solar values by \citet{lodders2003}
and the other abundances with respect to iron. Emission measures ($Y = \int n_{\mathrm{e}} n_{\mathrm{H}} dV$) 
are given in $10^{64}$ cm$^{-3}$.}
\label{tab:simulfit}
\begin{center}
\begin{tabular}{l|ccc}
\hline\hline
Parameter 	& 2T-model	& {\it wdem}-model	 \\
\hline
Y$_1$ 		& $0.86\pm0.03$		&			 \\
Y$_2$ 		& $9.97\pm0.13$		&			 \\
Y$_{\mathrm{wdem}}$&      	& $11.56\pm0.17$		 \\
$kT_1$ (keV)    & $0.80\pm0.01$		& 			 \\  
$kT_2$ (keV)	& $1.67\pm0.02$		&			 \\
$kT_{\mathrm{max}}$ & 		& $2.38\pm0.05$		 \\
$\alpha$	& 		& $0.47\pm0.02$	 	\\
c		& 		& $0.22\pm0.01$ 		\\
$kT_{\mathrm{mean}}$ &    	& $1.82\pm0.04$	 		\\
C/Fe		&  $0.73\pm0.11$& $0.74\pm0.13$ 		\\ 
N/Fe		&  $1.62\pm0.19$& $1.62\pm0.21$ 		\\  
O/Fe 		&  $0.60\pm0.03$& $0.59\pm0.04$ 		\\
Ne/Fe		&  $1.40\pm0.11$& $1.25\pm0.12$ 		\\
Mg/Fe		&  $0.70\pm0.05$& $0.60\pm0.06$ 		\\
Fe		&  $0.98\pm0.02$& $1.06\pm0.03$ 		\\
\hline
$\chi^2$ / $\nu$ &  2737/1905 		&  2728/1905 	\\
\hline
\end{tabular}
\end{center}
\end{table}

Since the difference of the position angles of the two observations is $187.3^{\circ}$, the stripes used to extract the spectra from the RGS are rotated with respect to each other by only $7.3^{\circ}$. While in the first observation the core of the emission falls on the center of the RGS detector, in the second observation the core of the emission is offset from the center of the detector by $\sim$1.5$\arcmin$. In order to obtain a combined spectrum with the highest possible statistics of the core of the galaxy, we extract spectra from a $1.1\arcmin$ wide stripe centred on the core of M~87 from both observations. We fit these spectra simultaneously. The fit to the total spectrum is shown in Fig.~\ref{total}.

We try two different thermal models to fit the emission of the hot gas in M~87: a combination of
two thermal components and a differential emission measure model (see section \ref{models}). The results of these two fits are shown in Table~\ref{tab:simulfit}. The reduced $\chi^{2}$s of both fits are $\sim1.4$, non of the thermal models can be ruled out by our analysis. 

Since the continuum is not very well determined by the RGS, absolute abundance values obtained by fitting different thermal models are slightly different. However, the relative abundances are consistent for both thermal models. In Table~\ref{tab:simulfit} we show our best fit abundances relative to iron, which is, due to the strong iron lines in the spectrum, the element with the best determined abundance value. The excellent statistics of the combined spectrum allows us, for the first time, to determine relatively accurate abundance values also for carbon and nitrogen. All the detected spectral lines are shown in Fig.~\ref{lines}, which shows the residuals of the RGS spectrum with line emission set to zero in the model. 

We note that \citet{sakelliou2002} use a cooling-flow model, with $T_{\mathrm{min}}=0.6$~keV, in combination with a single temperature model. By fitting this model we do not obtain a
better fit than with the 2-temperature model and the determined abundances (also the absolute abundances) are consistent with those obtained with the 2-temperature model. The abundance values determined by \citet{sakelliou2002} are different from our values due to a different power-law slope and normalisation used to account for the AGN emission. They obtained different parameter values for the power-law because they used a different (obtained by a survey at lower spatial resolution) Galactic absorption column density of $2.5\times10^{20}$~cm$^{-2}$.

The spectra lack emission from lines that trace gas cooling below 0.8~keV (e.g. O~VII line emission), which confirms the previous results of \citet{sakelliou2002}. The spectra, however, allow us to put strong upper limits on the amount of the cooling gas. Our upper limit on the emission measure of the gas with a temperature of $\sim0.2$~keV with the same oxygen abundance as the ambient hot gas is $Y=6.9\times10^{61}$~cm$^{-3}$. If
we assume that all this gas is in a sphere with a radius of $0.55\arcmin$ then our upper limit of the total mass of the $\sim0.2$~keV gas is $\approx10^7$~M$_{\odot}$, which is $\approx8$\% of the total gas in the sphere  ($\approx1.2\times10^{8}$~M$_{\odot}$).

\subsection{Intrinsic absorption or resonant scattering}

In order to detect intrinsic absorption by the putative cooling gas in the core of M~87, and/or by the warm-hot intergalactic medium associated with the Virgo cluster \citep[reported by][]{fujimoto2004}, we subtract the first observation from the second one. This way we get rid of most of the cluster emission, which allows us to analyze the excess power-law emission from the central AGN detected during the second observation. We do not see any absorption lines in the power-law spectrum. However, for the \ion{O}{vii} absorption column we find an upper limit of $10^{16}$~cm$^{-2}$. 

Resonant scattering of line emission is expected to be important in the dense cores of clusters of galaxies \citep[e.g.][]{gilfanov1987}. Its detection in the RGS spectra of the elliptical galaxy NGC~4636 in the Virgo cluster was reported by \citet{xu2002}. The best lines to detect resonant scattering in M~87 are the $2p-3s$ and the $2p-3d$ \ion{Fe}{xvii} lines at 17.1~\AA\ and 15.01~\AA\ respectively. The oscillator strength of the $2p-3d$ transition is substantially higher than that of the $2p-3s$ transitions. Their ratios at different distances from the core of the galaxy can reveal resonant scattering of the line at 15.01~\AA. Since these lines originate from the same ion of the same element, the observed spatial dependence of their ratios cannot be due to gradients in temperature or elemental abundances in the ISM. However, these lines can be well determined only close to the core of the galaxy. We determine the ratios of their equivalent widths in a 30\arcsec\ wide stripe going through the core of the galaxy and in the combined spectrum extracted from two 15\arcsec\ stripes next to the central extraction region.  While the ratio of the equivalent widths of the $2p-3s$ lines to that of the $2p-3d$ line in the central region is $1.3\pm0.3$, the ratio in the neighbouring bins is $0.7\pm0.5$. This result indicates that 15.01~\AA\ photons emitted in the core of the galaxy get scattered before they exit the interstellar medium. However, the large errors do not allow us to make strong conclusions.

\subsection{Radial profiles}

\begin{figure}
\begin{center}
\includegraphics[width=6.5cm,clip=t,angle=0.]{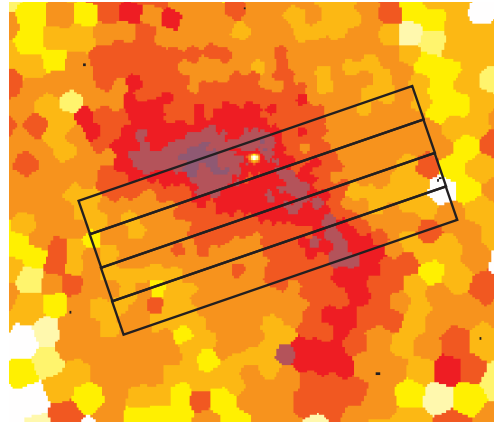}
\end{center}
\caption{The RGS extraction regions used for fitting the radial profiles overplotted on a temperature map by 
Aurora Simionescu (in preparation). Our radial profiles go along the South-Western lobe filled with cooler gas - clearly seen on the temperature map.} 
\label{strips}
\end{figure}

\begin{figure}
\includegraphics[width=6.5cm,clip=t,angle=270.]{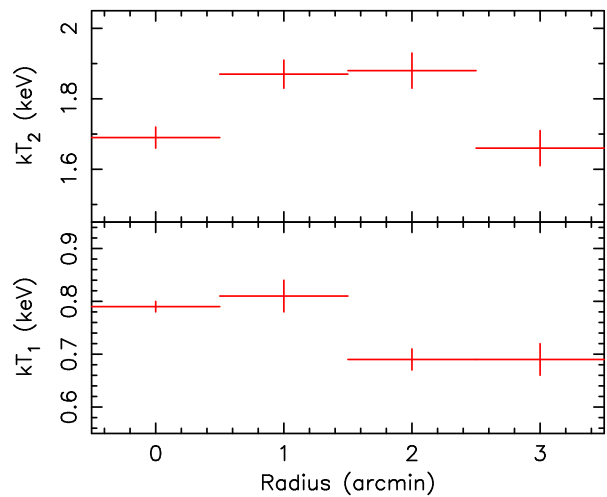}
\caption{Radial temperature profiles of the two thermal components obtained by fitting spectra extracted from rectangular regions in the cross-dispersion direction of the RGS.} 
\label{tempprof}
\end{figure}

\begin{figure}
\includegraphics[width=6.5cm,clip=t,angle=270.]{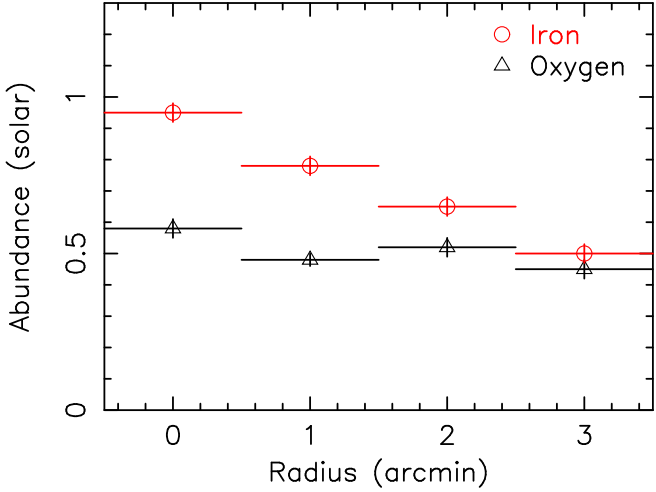}
\caption{Radial abundance profiles for oxygen and iron obtained by fitting spectra extracted from rectangular regions in the cross-dispersion direction of the RGS.} 
\label{abunprof}
\end{figure}

\begin{table}
\caption{Fit results for spatially resolved RGS spectra. We model the plasma with two thermal components. 
Abundances are given with respect to the proto-solar values by \citet{lodders2003}. The emission measures are given in $10^{64}$~cm$^{-3}$.}
\begin{center}
\begin{tabular}{l|cccc}
\hline\hline
Par.		 & -0.5 / 0.5$\arcmin$  & 0.5 / 1.5$\arcmin$  & 1.5 / 2.5$\arcmin$ & 2.5 / 3.5$\arcmin$  \\
\hline
$Y_1$			 &  $0.82\pm 0.04$ &  $0.23\pm0.03$  &  $0.16\pm0.02$  &  $0.18\pm0.02$  \\
$Y_2$			 &  $10.83\pm0.17$ &  $8.38\pm0.08$  &  $6.18\pm0.10$  &  $4.79\pm0.07$  \\
$kT_1$     		 &  $0.79\pm 0.01$ &  $0.81\pm0.03$  &  $0.69\pm0.02$  &  $0.69\pm0.03$  \\
$kT_2$ 			 &  $1.69\pm 0.03$ &  $1.87\pm0.04$  &  $1.88\pm0.05$  &  $1.66\pm0.05$  \\
C                        &  $0.63\pm 0.16$ &  $0.44\pm0.13$  &  $0.30\pm0.16$  &  $<0.18$  \\
N                        &  $1.64\pm 0.24$ &  $0.62\pm0.18$  &  $0.67\pm0.22$  &  $<0.23$	\\
O			 &  $0.58\pm 0.03$ &  $0.48\pm0.02$  &  $0.52\pm0.03$  &  $0.45\pm0.03$  \\
Ne			 &  $1.41\pm 0.12$ &  $1.17\pm0.13$  &  $0.89\pm0.17$  &  $<0.23$	\\
Fe			 &  $0.95\pm 0.03$ &  $0.78\pm0.03$  &  $0.65\pm0.03$  &  $0.50\pm0.03$  \\
Scale $s$		 &  $2.14\pm 0.10$ &  $1.10\pm0.05$  &  $1.18\pm0.09$  &  $1.42\pm0.08$  \\
\hline
$\chi^2$ / $\nu$	 &  1308/918    &  1265/918   &  1199/918  & 1187/918	 \\
\hline
\end{tabular}
\label{tabprof}
\end{center}
\end{table}

The excellent statistics of the RGS spectra obtained during the second observation allows us to extract spectra from four $1\arcmin$ wide strips in the cross-dispersion
direction of the instrument and make radial temperature and abundance profiles up to $3.5\arcmin$ from the core of the galaxy. Unfortunately the core of the emission is offset
from the center of the RGS detector so we could not determine the profiles from both sides of the core. 

Since the 2-temperature model describes the data equally well as the {\it{wdem}} model and because the 2-temperature model may also be well justified by two separate thermal
components within and outside of the radio lobes in M~87, we fit all the spectra with the more simple 2-temperature model. 
The best fit results are shown in Table~\ref{tabprof}.
We note that the extraction regions from which we determine the RGS radial profiles are actually sampling the South-Western lobe at different distances from the core (see Fig.~\ref{strips}).

The radial temperature profiles for the two thermal components are shown in Fig.~\ref{tempprof}. They both show an increase of temperature in the $0.5-1.5\arcmin$ region and
they do not show a radial gradient. The radial distribution of the emission measures shows a $\sim3$ times larger relative contribution of the cooler gas in the central region than
in the $0.5-2.5\arcmin$ region. 

The radial abundance profiles for oxygen and iron are
shown in Fig.~\ref{abunprof}. We see that the radial distributions of oxygen and iron, the two best
determined abundances, are different. Both elements have a peak in the core of the cluster, but while the oxygen abundance drops at $1\arcmin$ distance from the core by
$\sim$17\%\ and then its distribution stays flat, the distribution of iron shows a gradient all the way out in our
investigated region. 

The radial distribution of neon also seems to show a gradient throughout the investigated region. But its abundance values are more uncertain. In the extraction regions not going through the core of the cluster the spatial extent of the source along the dispersion direction broadens the lines in the RGS and the neon line gets blended with the iron lines. Thus, for extraction regions further away from the core of the galaxy, the systematic uncertainties on the neon abundance get higher. 

The radial distribution of carbon and nitrogen seems to be consistent with that of iron, however their values are too uncertain to draw conclusions.

We note that the best fit temperature and abundance values determined from the RGS are actually the mean values along the dispersion direction of the instrument and the true radial profiles have stronger gradients than those derived from the RGS spectra.

\section{Discussion}
\label{discussion}

\subsection{Chemical enrichment by type Ia and core collapse supernovae}
\label{discus:enrichment}

\begin{table}
\caption{Comparison of the observed mass ratios in the core of M~87 for five elements to iron with the predicted mass ratios assuming that all the elements were produced only by supernovae with 86\% of all iron produced by SN~Ia. The observed to predicted ratio for oxygen is here by definition one.}
\begin{center}
\begin{tabular}{l|ccc}
\hline
\hline
Mass ratios & Obs.   &	Pred.  & Obs./Pred. \\
\hline
M(C)/M(Fe)  & 1.32 &  0.13  & 10  \\
M(N)/M(Fe)  & 0.94 &  0.003 & 313 \\
M(O)/M(Fe)  & 2.89 &  2.89  & $\equiv1$    \\
M(Ne)/M(Fe) & 1.32 &  0.36  & 3.7   \\
M(Mg)/M(Fe) & 0.37 &  0.19  & 1.9   \\
\hline
\end{tabular}
\label{PredObs}
\end{center}
\end{table}

The ISM/ICM in M~87 was enriched by heavy elements primarily by core collapse supernovae (SN$_{\mathrm{CC}}$), type Ia supernovae (SN~Ia) and stellar winds. 

In order to investigate the relative contribution of SN~Ia and SN$_{\mathrm{CC}}$ one can use the observed ratios of elements (produced by supernovae) with detected line emission to iron to find the best fit to the following function:

\begin{equation}
\left( \frac{M(\mathrm{X})}{M(\mathrm{Fe})}
\right)_{\mathrm{cluster}}=f\left(\frac{M(\mathrm{X})}{M(\mathrm{Fe})}\right)_{\mathrm{SNIa}}+(1-f)\left(\frac{M(\mathrm{X})}{M(\mathrm{Fe})}\right)_{\mathrm{SN_{CC}}},
\end{equation}

\noindent where the expression $M(\mathrm{X})/M(\mathrm{Fe})$ represents the fraction of the mass of a given element to the mass of iron as observed in the galaxy (cluster), as predicted for the
yield of SN~Ia and as predicted for the yield of SN$_{\mathrm{CC}}$; $f$ represents the fraction of iron produced by SN~Ia relative to the total mass of iron produced by the two main types of supernovae.  

For nucleosynthesis products of SN$_{\mathrm{CC}}$ we adopt an average yield of stars on a mass range from 10~M$_\odot$ to 50~M$_\odot$ calculated by \citet{tsujimoto1995}
assuming a Salpeter initial mass function. For nucleosynthesis products of SN~Ia we adopt values calculated for the delayed-detonation model WDD2 by \citet{iwamoto1999}.

The best determined are the oxygen and iron abundance. Assuming that all iron and oxygen is produced by supernovae, from their ratios we estimate that 86\% of iron is coming
from SN~Ia. Then the relative number of SN~Ia contributing to the enrichment of the ICM is $\sim40$\% while the relative number of SN$_{\mathrm{CC}}$ is $\sim60$\%. We note that \citet{bohringer2005} based on the XMM-Newton EPIC measurements analysed by \citet{matsushita2003} estimate that about 80\% of iron comes from SN~Ia. Given the different extraction regions and the gradient in the iron abundance, this estimate is in a good agreement with our result. The measured neon abundance is not matched well by the supernova models. It is much higher than what the models predict (see Table~\ref{PredObs}), which was previously also observed in the cluster of galaxies 2A~0335+096 \citep{werner2006}. The best fit of the magnesium abundance value may be influenced by the uncertainties in the effective area calibration of the RGS at short wavelengths. Therefore, we do not discuss it here.

The supernovae produce only small amounts of carbon and nitrogen.
As shown in Table~\ref{PredObs}, we observe about 300 times more nitrogen and 10 times more carbon in M~87 than what can be explained by the employed supernova models. 
The difference between the predicted mass produced by all supernovae (as calculated from the total mass of iron, relative numbers of SN~Ia and SN$_{\mathrm{CC}}$ and the theoretical yields) and the total observed mass of these elements can be used to put an upper limit on the contribution from stellar winds to the chemical enrichment of the ICM. 

However, to estimate the total mass of these metals is difficult. The long and narrow extraction region of RGS and the disturbed nature of the core allow us only to make rough estimates. If we assume that the measured central abundances are typical for the inner $2.5\arcmin$ of the galaxy and we consider an average density of $2.0\times10^{-2}$~cm$^{-3}$ \citep[based on][]{matsushita2002}, than the total mass of carbon and nitrogen in the ISM is $\approx7\times10{^6}$~M$_{\odot}$ and $\approx5\times10^{6}$~M$_{\odot}$ respectively. However, the radial abundance profiles of nitrogen and carbon suggest abundance gradients, which means that since the measured abundance values are an average along the dispersion direction of the RGS, the true central abundance is probably higher. Because of the assumptions, the estimated masses can be considered as lower limits.

\subsection{Carbon and nitrogen abundances and enrichment by AGB stars}

\begin{figure}[tb]
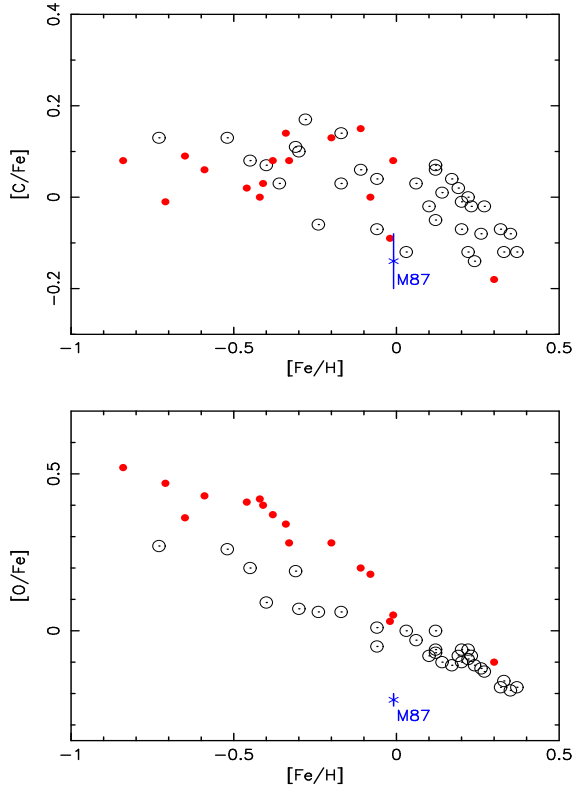

\begin{minipage}{7.5cm}
\begin{center}
\includegraphics[width=5cm,height=7.5cm,clip=t,angle=270.]{fig6a.ps}\\
\end{center}
\end{minipage}
\begin{minipage}{7.5cm}
\begin{center}
\includegraphics[width=5cm,height=7.5cm,clip=t,angle=270.]{fig6b.ps}
\end{center}
\end{minipage}
\caption{Comparison of the [C/Fe] and [O/Fe] enrichment in M~87 with those
of the Galactic thin and thick disk stars of \citet{bensby2006}. Thin and thick disc stars are marked by open and filled circles respectively. The value for M~87 is indicated by the asterisk. } 
\label{bensby}
\end{figure}

\begin{figure}[tb]
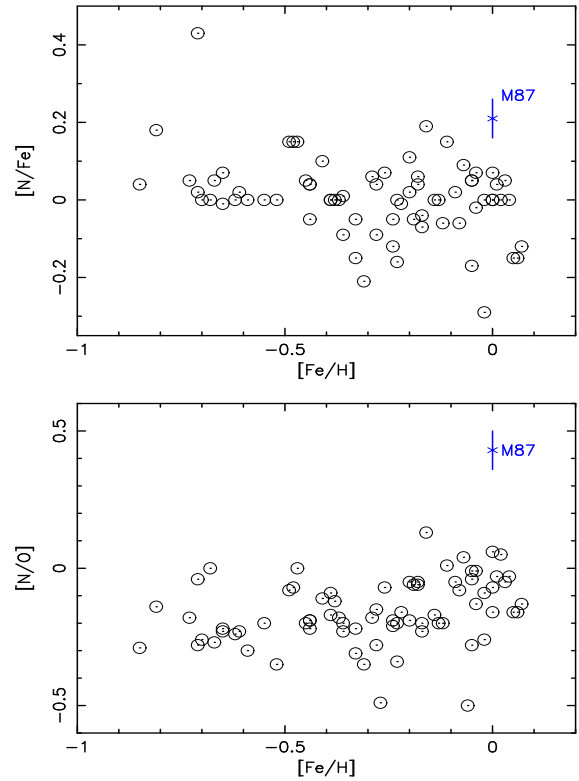

\begin{minipage}{7.5cm}
\vspace{0.5cm}
\begin{center}
\includegraphics[width=5cm,height=7.5cm,clip=t,angle=270.]{fig7a.ps}
\end{center}
\end{minipage}
\begin{minipage}{7.5cm}
\begin{center}
\vspace{0.25cm}
\includegraphics[width=5cm,height=7.5cm,clip=t,angle=270.]{fig7b.ps}
\end{center}
\end{minipage}
\caption{Comparison of the [N/Fe] and [N/O] enrichment in M~87 with the Galactic stellar population of \citet{chen2000} and \citet{shi2002}. The value for M~87 is indicated by the asterisk.} 
\label{Shi}
\end{figure}

While it is well known that elements from oxygen up to the iron group are primarily produced in supernovae, the main sites of the carbon and nitrogen production are still being debated. It is believed that both elements are being contributed by a wide range of sources. The main question is whether their production is dominated by the winds of short-lived massive stars or by the longer-lived progenitors of AGB stars. 

An analysis of the enrichment history of carbon in our Galaxy has for example been performed 
by \citet{bensby2006}. They find that the 
ratio C/Fe is fairly constant with a decrease of
about a factor of 1.5 over the chemical evolution
history involving [Fe/H] values from -1 to +0.4
as covered in their survey. At the same time the 
O/Fe ratio decreases by about a factor of 5. The general
explanation of this large decrease in the relative oxygen 
abundance is an early enrichment by SN$_{\mathrm{CC}}$ with 
high O/Fe values and a subsequent primarily iron production
by SN~Ia, which go off on much longer time scales than the SN$_{\mathrm{CC}}$. 
In contrast the flat C/Fe ratio then implies
that the enrichment by Fe through SN Ia and the C pollution
has occurred very much in parallel. The similar enrichment time-scale
suggests that the main sites of carbon production are the longer-lived progenitors of AGB stars.  

For our case of the, on average, much older stellar population 
of M~87 it is now interesting to compare the enrichment 
situation with that of our Galaxy. In Fig.~\ref{bensby} we compare
the C/Fe and O/Fe enrichment in M~87 with those
of the galactic thin and thick disk stars of \citet{bensby2006}. 
We see that while O/Fe is significantly lower in the ISM of M~87 
than in the stellar population of our Galaxy, C/Fe is comparable to 
the lowest C/Fe values detected in the Milky Way. The O/Fe ratio in M~87
is so low probably due to the fact that the contribution of SN~Ia to the enrichment of the ISM in M~87 is approximately 3 times as large as in our Galaxy (according to \citet{tsujimoto1995} the relative number of SN~Ia for our Galaxy is $N_{\mathrm{Ia/Ia+II}}=0.13$). The low O/Fe ratio also suggests that the star formation in M~87 essentially stopped. 
The C/Fe ratio is consistent with the Galactic value which indicates that the enrichment by Fe from SN~Ia and by C through AGB stars is occurring also in M~87 in parallel.

In Fig.~\ref{Shi} we compare the N/Fe and N/O enrichment with a sample of Galactic stars published by \citet{chen2000} and \citet{shi2002}. The measured N/O ratio in M~87 is higher than in the sample of Galactic stars. However, the N/Fe ratio is comparable to the highest values detected in this sample. Given the old stellar population in M~87 the high N/O ratio in the ISM is not surprising. While the oxygen (and probably also some of the nitrogen) was supplied by core collapse supernovae early in the enrichment history, nitrogen is still being constantly supplied by intermediate mass AGB stars on a time scale possibly similar to the enrichment by iron.

\subsection{Spatial abundance distribution}

While the chemical abundances of all elements have a peak in the center of the galaxy, the radial abundance profiles indicate that the spatial distributions of oxygen and iron are different. The radial distribution of iron, primarily a type Ia supernova product, has a gradient throughout the field of view of the RGS. However, the radial distribution of oxygen, primarily a product of core collapse supernovae, seems to be flat with a projected abundance of $\sim0.5$ solar, with only a modest peak in the extraction region centred on the core of the galaxy. 
While the flat component of the distribution of core collapse supernova products has been produced by
supernovae going off primarily in the earlier stages of the evolution of the galaxy, type Ia supernova still continue to enrich the ISM/ICM in M~87. 

Our results confirm the previous results of \citet{bohringer2001}, \citet{finoguenov2002} and \citet{matsushita2003} who based on observations with EPIC showed that oxygen has a relatively flat radial abundance distribution compared to the steep gradients of the heavier elements. While the low spectral resolution of EPIC combined with the relative closeness of the iron lines and calibration problems at low energies left some space for doubts about these early results, the RGS with its high spectral resolution showed conclusively that the spatial distributions of oxygen and iron are different.
However, the systematic uncertainties in the absolute abundance values leave still open the possibility of more enhanced abundance values in the innermost bin fitted with the RGS, which would mean a stronger central peak in the distribution of iron, but also a peak in the oxygen distribution.

\section{Conclusions}

We have analysed the temperature structure and the chemical abundances in the giant elliptical galaxy M~87 using high-resolution spectra obtained during two deep XMM-Newton observations  with the Reflection Grating Spectrometers.
We found that:

\begin{itemize}
\item
While we confirm the two-temperature structure of the ISM we show that a multi-temperature {\it{wdem}} fit also describes the data well and gives the same relative abundances as the two-temperature fit.  
\item
The O/Fe and O/N ratios in the ISM of M~87 are lower than in the stellar population of our Galaxy, which shows that the relative contribution of core collapse supernovae in the old stellar population of M~87 to the enrichment of the ISM was significantly less than in our Galaxy.
\item
The comparison of the C/Fe and N/Fe ratios in the ISM of M~87 with those in the stellar population of our Galaxy suggests that the enrichment of the ISM by iron through Type~Ia supernovae and by carbon and nitrogen is occurring in parallel and thus the dominant source of carbon and nitrogen in M~87 are the intermediate- and low-mass AGB stars. 
\item
From the oxygen to iron abundance ratio we estimate that the relative number of core collapse and type Ia supernovae contributing to the enrichment of the ISM in the core of M~87 is $\sim60$\% and $\sim40$\% respectively.
\item
The spatial distributions of oxygen and iron in M~87 are different. While the oxygen abundance distribution is flat the iron abundance peaks in the core and has a gradient throughout the field of view of the instrument, suggesting an early enrichment by core-collapse supernovae and a continuous contribution of SN~Ia.
\end{itemize}

\begin{acknowledgements}
This work is based on observations obtained with XMM-Newton, an ESA science mission with instruments
and contributions directly funded by ESA member states and the USA (NASA). The Netherlands Institute
for Space Research (SRON) is supported financially by NWO, the Netherlands Organization for Scientific Research. 
We would like to thank Onno Pols for reading and giving constructive comments on draft version of the paper. 
\end{acknowledgements}

\bibliographystyle{aa}
\bibliography{bib}

\end{document}